\documentclass[manuscript]{aastex}

\newcommand{\etal}{{\it et al.~}}

\begin{document}

\title{Weak Lensing Detection of Cl 1604+4304 at $z = 0.90$}

\author{V. E. Margoniner\altaffilmark{1}}

\author{L. M. Lubin\altaffilmark{1}}

\author{D. M. Wittman\altaffilmark{1}}

\author{G. K. Squires\altaffilmark{2}}

\altaffiltext{1}{Department of Physics, University of California at Davis, One Shields Avenue, Davis, CA 95616}
\altaffiltext{2}{Spitzer Science Center, California Institute of Technology, Pasadena, CA 91125}

\begin{abstract}

We present a weak lensing analysis of the high-redshift cluster Cl
1604+4304.  At $z=0.90$, this is the highest-redshift cluster yet
detected with weak lensing.  It is also one of a sample of
high-redshift, optically-selected clusters whose X-ray temperatures
are lower than expected based on their velocity dispersions.  Both the
gas temperature and galaxy velocity dispersion are proxies for its
mass, which can be determined more directly by a lensing analysis.
Modeling the cluster as a singular isothermal sphere, we find that the
mass contained within projected radius $R$ is $3.69\pm1.47 \times
({R\over 500 {\rm kpc}}) 10^{14}$ $M_\odot$.  This corresponds to an
inferred velocity dispersion of $1004\pm199$ km s$^{-1}$, which agrees
well with the measured velocity dispersion of $989^{+98}_{-76}$ km
s$^{-1}$ \citep{gal04}. These numbers are higher than the
$575^{+110}_{-85}$ km s$^{-1}$ inferred from Cl 1604+4304 X-ray
temperature, however all three velocity dispersion estimates are
consistent within $\sim 1.9\sigma$.

\end{abstract}


\section{Introduction}

Clusters of galaxies have historically been detected through their
baryon content, either in optical/near-infrared surveys searching for
overdensities of galaxies or in X-ray surveys searching for emission
from the hot, intracluster medium. Previous observations have found
strong differences between X-ray and optically-selected clusters at
moderate-to-high redshift, implying that at least some massive
clusters do not obey the local X-ray--optical relations.  Specifically,
the properties of the galaxies and the intracluster gas in local Abell
clusters are strongly related. There exist well-defined correlations
between the X-ray properties of the gas, such as luminosity ($L_x)$
and temperature ($T_x$), and the optical properties of the galaxies,
such as blue luminosity ($L_B$) and velocity dispersion
($\sigma$). These relations indicate that the galaxies and gas are in
thermal equilibrium, i.e.\ $T_x \propto \sigma^2$ (Edge \& Stewart
1991).  Moderate-redshift clusters up to $z \sim 0.5$ exhibit the same
X-ray--optical relations (Mushotzky \& Scharf 1997; Horner 2001).

However, X-ray observations of all optically-selected clusters at 
$z\gtrsim 0.5$ indicate that they are weak X-ray sources, regardless of
their measured richness or velocity dispersion, with luminosities of
only a few $\times 10^{44}$ ergs s$^{-1}$ (Castander et al.\ 1994;
Bower et al.\ 1997; Holden et al.\ 1997; Donahue et al.\ 2001; Lubin,
Oke, \& Postman 2002; Lubin, Mulchaey \& Postman 2004). Consequently,
many optically-selected clusters do not obey the local $L_x-\sigma$
relation. Their X-ray luminosities are low for their velocity
dispersions, indicating that optically-selected clusters at these
redshifts are underluminous (by a factor of $3-40$) compared to their
X-ray--selected counterparts. These results seem to imply that, at
least for some clusters, the galaxies and gas are no longer in thermal
equilibrium and that the clusters are still dynamically young.  Strong
observational evidence, such as double peaks, significant
substructure, and/or a filamentary appearance in the gas, galaxy, and
total mass distributions, does indicate that many (if not most)
clusters at these redshifts are actively forming (see e.g., Lubin \&
Postman 1996; Henry et al.\ 1997; Gioia et al.\ 1999; Della Ceca et
al.\ 2000; Ebeling et al.\ 2000; Jeltema et al.\ 2001; Stanford et
al.\ 2001; Hashimoto et al.\ 2002; Maughan et al.\ 2003; Huo et al.\
2004)

Cl 1604+4304 at $z = 0.90$, orginally detected in the Gunn, Hoessel \&
Oke (1986) survey, is typical of optically-selected clusters at these
redshifts. Velocity dispersions measured first by Postman et al.\
(2001) and more accurately by Gal \& Lubin (2004) indicate that the
system is massive, equivalent to an Abell richness class 1--2
cluster. It has, however, only a modest X-ray luminosity and
temperature of $L_x = 2.0 \times 10^{44}$ h$_{70}^{-2}$ ergs s$^{-1}$
and $T_x = 2.51^{+1.05}_{-0.69}$ keV, respectively (Lubin et al.\
2004). Based on its measured velocity dispersion of $989^{+98}_{-76}$
km s$^{-1}$, Cl 1604+4304 is cooler by a factor of 2--4 compared to
the local $\sigma-T_x$ relation.  

Gas temperature and galaxy velocity dispersion are only proxies for
mass, which can be determined directly by a lensing analysis.  The
mass determined by lensing is independent of the cluster's dynamical
state, star formation history, and baryon content (Smail, Ellis \&
Fitchett 1994; Smail \etal\ 1995).  As a result, we have undertaken a
weak lensing analysis of Cl 1604+4304 using deep imaging from the Keck
10-m telescope. Because weak lensing uses the background galaxy
population, it becomes progressively more difficult to apply to
high-redshift clusters. The highest-redshift cluster previously
detected with lensing is MS 1054-03, a very X-ray luminous, hot,
massive cluster at $z=0.83$ (Luppino \& Kaiser 1998; Hoekstra, Franx,
\& Kuijken 2000).  Its X-ray luminosity is $1.6 \times 10^{45}$
h$_{70}^{-2}$ ergs s$^{-1}$, and estimates of its X-ray temperature
range from 7.2$^{+0.7}_{-0.6}$ keV to 12.3$^{+3.1}_{-2.2}$ keV (Gioia
\etal\ 2004 and references therein). Hoekstra \etal\ (2000) determined
MS 1054-03 weak lensing mass to be $(1.2\pm0.2)\times10^{15}$ within 1
Mpc radius, corresponding to a velocity dispersion of
$1311^{+83}_{-89}$ km s$^{-1}$ for an isothermal sphere, in very good
agreement with the spectroscopicaly measure of $1153\pm80$ km s$^{-1}$
(Gioia \etal\ 2004; Tran \etal\ in preparation).

Although Cl 1604+4304 has a substantially lower gas luminosity and
temperature than MS 1054-03, it may be more typical of galaxy clusters
at these redshifts because the number density of very X-ray luminous
clusters declines significantly at redshifts of $z > 0.6$ (e.g.,
Mullis et al.\ 2004).

In addition to deep imaging, weak lensing analyses of high-redshift
clusters requires better knowledge of the source redshift
distribution.  To calibrate the mass of a lens, one must know the mean
distance ratio of the sources.
For typical source redshifts around unity, uncertainties of the order
of 10\% in the source redshift distribution matter little for a
low-redshift cluster, but they represent a large systematic error when
the lensing cluster is itself at a redshift of $z \sim 1$. Here, the
tail of the source redshift distribution is behind the cluster, and it
becomes crucial to minimize the errors in estimating this tail.  In
this paper, we introduce a new method of estimating the source
redshift distribution based on degrading higher-quality data rather
than estimating the faint tail of one's own data. In \S 2 \& 3, we
describe the data and our weak lensing analysis. In \S 4, we summarize
our results and discuss future plans for a more accurate weak lensing
measurement of this cluster.

\section{The Data}

The imaging used in the weak lensing analysis was obtained on 27 July
1999 at the Keck 10-m telescope using the Low Resolution Imaging
Spectrograph (LRIS; Oke et al.\ 1986). The pixel scale of LRIS is
$0.215^{\prime\prime}$, and its field of view is approximately
$5.0^{\prime}\times6.5^{\prime}$. The seeing was excellent at
$0.60^{\prime\prime}$ full-width-half-maximum (FWHM) in the center of
the image, with some degradation to an average of
$0.75^{\prime\prime}$ at the edges.  Twenty-nine $R$-band images with
exposure times of 300 sec each were taken. A dither pattern of
$40^{\prime\prime}$ in width was used to minimize the effect of bad
columns and bright stars. Conditions were not photometric; however,
photometric calibration was obtained from a shallower observation of
the same area, with the same instrument, on a photometric night.
These calibration images were taken as part of the Oke et al.\ (1998)
survey of nine high-redshift clusters of galaxies (see also Postman et
al.\ 1998, 2001).

\begin{figure}
\centering \includegraphics[width=.70\columnwidth]{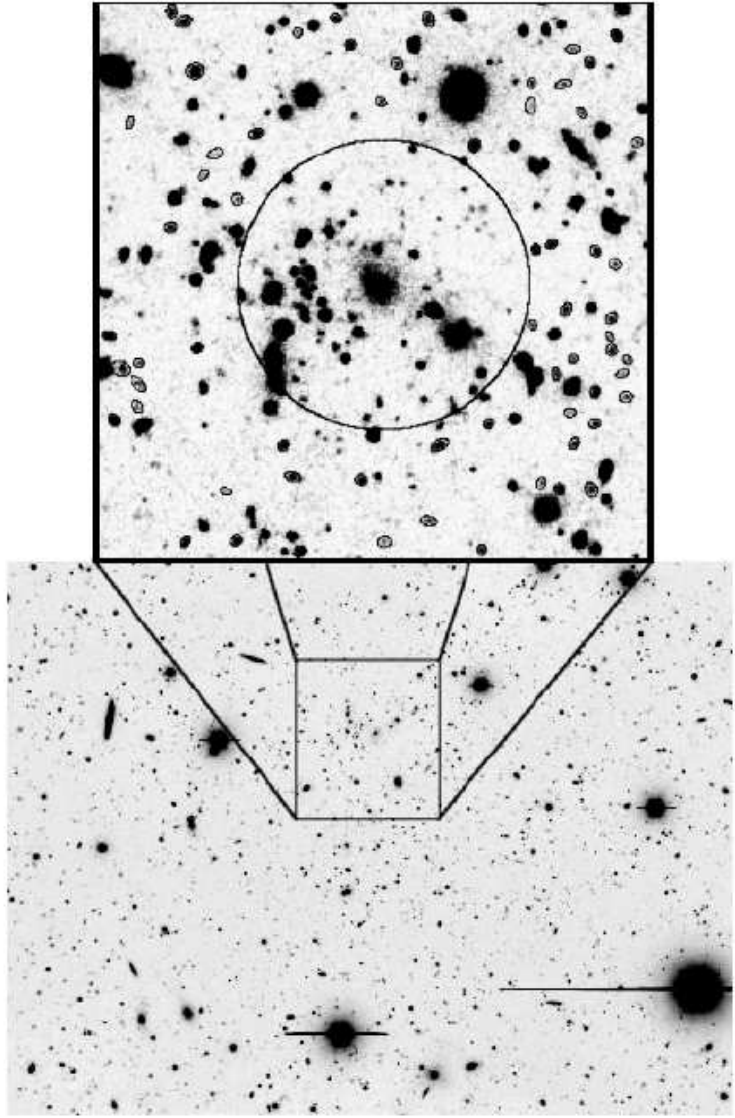}
\caption{Deep R imaging. Left panel: entire
$5.0^{\prime}\times6.5^{\prime}$ field of view. Right panel:
$1.5^{\prime}\times1.5^{\prime}$ zoom around the cluster optical
center. The circle indicates the inner cut in radius (objects too
close to the center were excluded from the lensing analysis), and
contours are shown for objects used in the weak lensing analysis (all
with $R>25$).}
\label{fig:image}
\end{figure}

\begin{figure}
\centering \includegraphics[width=.70\columnwidth]{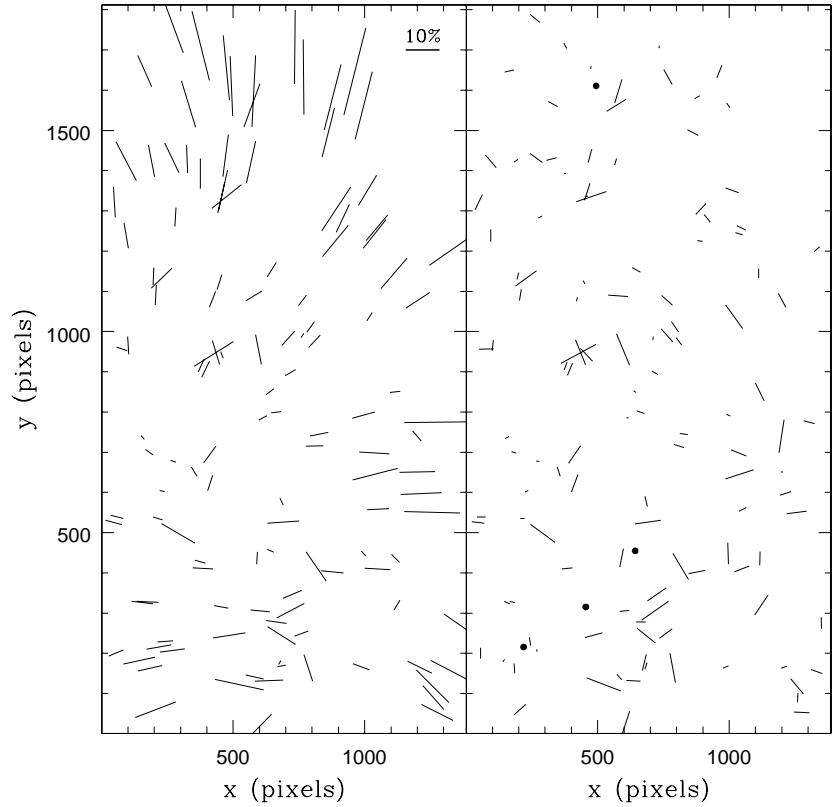}
\caption{Point-spread function correction.  Shapes of stars, which as
point sources should be perfectly round, are represented as sticks
encoding ellipticity and position angle.  Left panel: raw data with
spatially varying PSF ellipticities up to 23\% (mean of 9\%).  Right
panel: after convolution with a spatially varying asymmetric kernel,
ellipticities are vastly reduced (stars with $\epsilon<0.5\%$ are
shown as dots), and the residuals are not spatially correlated as a
lensing signal would be.}
\label{fig:circ}
\end{figure}

The imaging data was reduced using standard IRAF routines, with the
usual corrections (including bias and flatfielding) being
applied. Individual exposures were aligned and co-added to create a
single, deep $R$-band image with an effective exposure time of 8700
sec. The final image is shown in the left panel of
Figure~\ref{fig:image}. To perform our weak lensing analysis, the
intrinsic anisotropy of the point-spread function (PSF) has to be
corrected.  The left panel of Figure~\ref{fig:circ} shows the
ellipticity and position angles of 127 stars as a function of
position, before correction.  The mean ellipticity is $9\%$, with a
rms of $5\%$, and the position angles are highly spatially correlated.
We convolved the image with a kernel with ellipticity components
opposite to that of the PSF at each point \citep{ft}.  However, the
3$\times$3 pixel kernel used was not large enough to correct the PSF
completely in one pass.  Therefore we iterated four times, until
improvement stopped.  Each iteration also allowed us to clip a few
more interloping compact galaxies from the PSF sample, leaving a clean
sample of 127 stars at the end.  The right panel of
Figure~\ref{fig:circ} shows the result. The mean ellipticity decreased
to $4\%$ with a $3\%$ rms, and the position angles are no longer
spatially correlated.  After rounding, the final FWHM varied from
$0.65^{\prime\prime}$ at the center to an average of
$0.85^{\prime\prime}$ at the edges.


The object detection and photometry was performed with SExtractor
version 2.2.2 \citep{sex}, resulting in a catalog of 4424 objects (136
per arcmin$^2$). The galaxy counts peak at $R \approx 27$, and the
magntiude limit is $R = 29$ for a 5$\sigma$ detection.

\section{The Weak Lensing Analysis}

We measure weighted moments of objects using the {\tt ellipto}
software described in \citep{ellipto}, discarding any sources which
triggered error flags.  We also use their seeing correction procedure,
which corrects for the dilution of the shear signal due to the
isotropic smearing of the PSF.  We discarded sources which were not at
least 25\% larger than the PSF at that position, rather than apply a
large and noisy correction. Because of the strong stellar size
dependence with position on the CCD (up to 30\%), a second order
spatial fit was used in the the seeing correction.  This spatial
variation in the size selection function does not bias the shapes of
the objects, but it does impose a spatial variation on the source
redshift distribution, which will be discussed below.

Objects with ellipticity greater than 0.6 tend to be blends of
multiple sources and were also excluded from the lensing analysis.
The number of objects passing all these quality checks is 1588 (49 per
arcmin$^2$).

We measure the tangential shear around the cluster optical center,
using background galaxies with $R>25$ (see right panel of
Figure~\ref{fig:image}). Because the area closest to the core is
expected to be dominated by cluster members, we use only sources at
Radius $>25^{\prime\prime}$ for the weak lensing analysis. The final
number of galaxies is 1053.

\begin{figure}
\centering \includegraphics[width=.70\columnwidth]{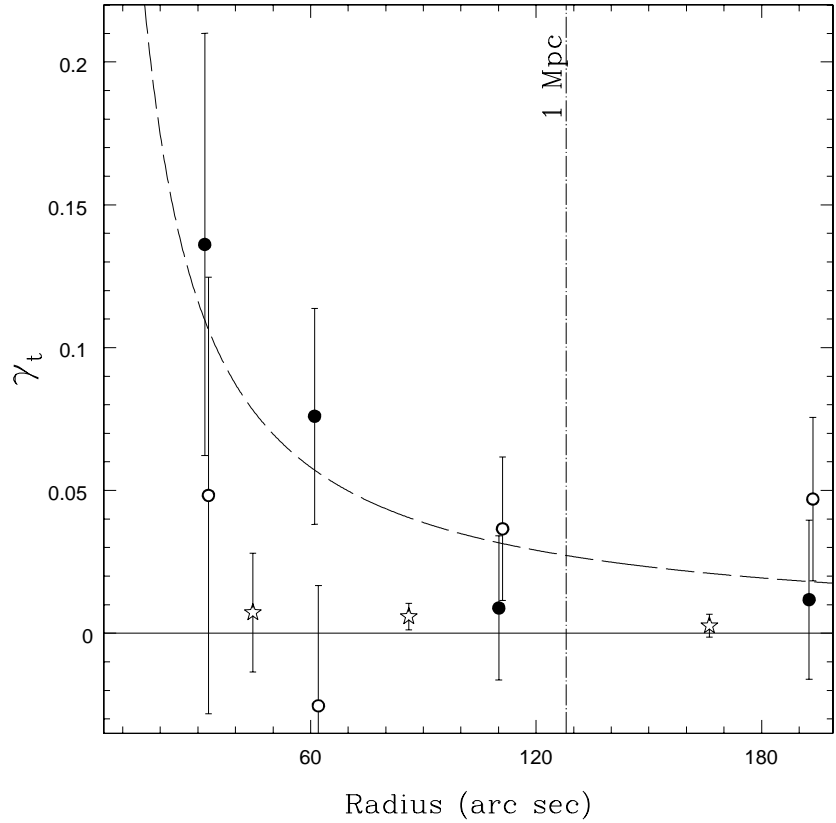}
\caption{Tangential shear, centered on the cluster optical center
(solid circles), as a function of radius, with 1$\sigma$ errors
shown. The dashed curve shows the best-fit isothermal sphere. The open
circles indicate the $45^\circ$ shear component, which should average
to zero. The open stars indicate the level of PSF systematics computed
from the stars (the radii are different because there are not enough
stars to subdivide the stars in four annulii). }
\label{fig:ann}
\end{figure}

Figure~\ref{fig:ann} shows the mean tangential shear for four annuli
around the optical center, with 1$\sigma$ errors attached.  As a null
test, we also show in open circles the other (45$^\circ$) shear
component, which is consistent with zero as expected. These
measurements are also indicated in Table~\ref{tab:shear}. The
tangential shear is detected at 1.8 and 2.0$\sigma$ respectively in
the two inner points, which has only a 2.5\% probablility of happening
randomly.  Hence the detection confidence is 97.5\%.  We do not expect
to detect shear in the outer two points, given the redshift and
velocity dispersion previously measured for this cluster (see below).
As a measurement of the expected level of PSF systematics, we also
show in Figure~\ref{fig:ann} that the ``tangential shear'' computed
from the stars is consistent with zero.

To avoid the mass sheet degeneracy inherent in mass reconstructions of
small fields, we fit a model to the shear profile. Given the low
signal to noise, we choose to fit the simplest possible model, a
singular isothermal sphere, to all four points. Assuming a cosmology
of $H_0 = 70$, $\Omega_m = 0.3$, and $\Omega_\Lambda = 0.7$, we find
that $\gamma_t = (0.054 \pm 0.021)\times({R\over{500~{\rm
kpc}}})^{-1}$.  The $\chi^2$ for this fit is 1.25 for three degrees of
freedom, hinting that the errors may have been overestimated.  As a
control, we fit an SIS to the 45-degree component of the shear and
found 0.027 $\pm$ 0.022, with a $\chi^2$ of 4.09 for three degrees of
freedom.  

To determine the mass, we must know the cluster redshift and the
source redshift distribution. For any given mass, the tangential shear
is proportional to the combination of {\it angular diameter distances}
from the observer to the lens ($D_{L}$), from the observer the source
($D_{S}$), and from the lens to the source ($D_{LS}$):

$$\gamma^{t} \propto {{D_{L}D_{LS}} \over {D_{S}}}$$

The cluster has a spectroscopic redshift of $z = 0.9001$
\citep{gal04}.  The challenge then is to estimate $\langle
{{D_{L}D_{LS}} \over {D_{S}}} \rangle$ for the source population.
This is not the same as estimating the mean redshift of the sources,
because the relation is quite nonlinear, and only sources with
redshifts of $z>0.9$ have a non-zero contribution.  For high-redshift
lenses, the estimation of the higher-redshift tail can be little more
than guesswork unless additional data are brought to bear.

We estimate the source redshift distribution by degrading the Hubble
Deep Field North (HDF-N), which has a well-known redshift
distribution, to match our data.  This is a more stable approach than
attempting to extrapolate the ditribution from our data alone, because
many of the background galaxies are near the faint limit of our
survey. First, we convolve the F606W (which is the closest match to
our filter) HDF-N image \citep{hdfn} with $0.65^{\prime\prime}$ and
$0.85^{\prime\prime}$ FWHM gaussian to simulate the range of PSF sizes
in the Keck image. Then, we re-pixelize, add noise, and catalog the
image to match the Keck data. We judge the match to be satisfactory
because the galaxy counts $N({\rm mag})$ in the degraded HDFN and Keck
catalogs are very similar.  

\begin{figure}
\centering \includegraphics[width=.70\columnwidth]{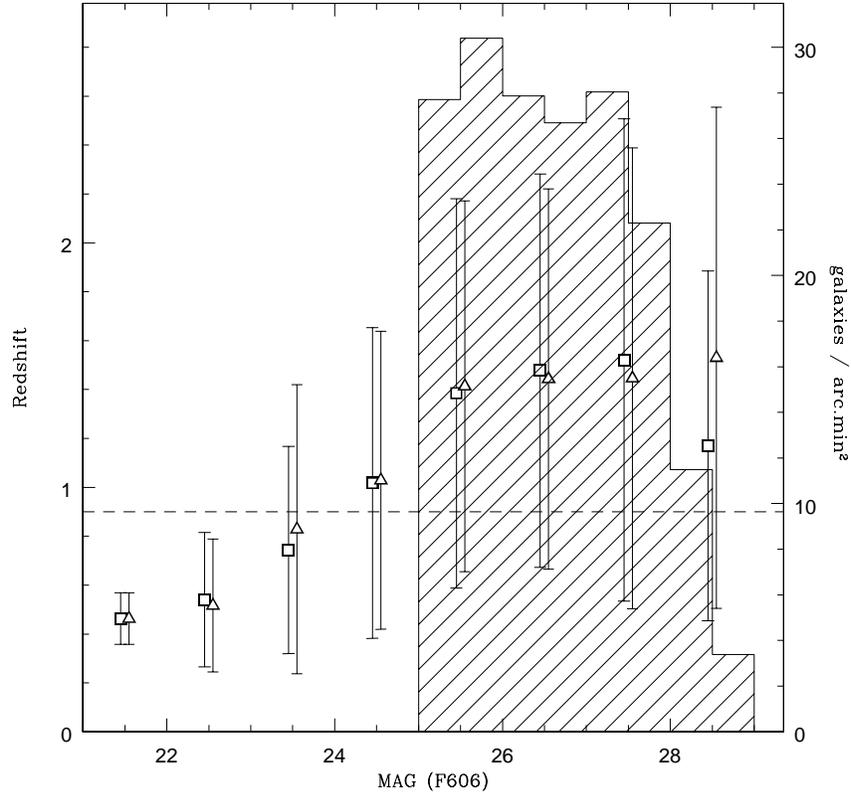}
\caption{Mean redshift and $1\sigma$ scatter as a function of
magnitude for HDFN simulations at $0.65^{\prime\prime}$ (squares) and
$0.85^{\prime\prime}$ (triangles) seeing. Mean redshifts are computed
discarding objects that would not be used in the lensing analysis. The
histogram indicates the distribution of magnitudes of lensing sources
in Cl1604.}
\label{fig:magz}
\end{figure}

We then apply the same magnitude and size cuts used in the lensing
analysis of the Keck image, and look up their photometric redshifts
\citep{fernandez-soto99}.  Figure~\ref{fig:magz} shows the mean
photometric redshift and $1\sigma$ scatter as a function of magnitude
for both PSF sizes.  The histogram indicates the distribution of
magnitudes of the lensing sources in Cl1604. Throughout the relevant
magnitude range, the mean source redshift is nearly constant at
$z=1.45$.  We convert each source photometric redshift to a distance
ratios ${{D_{L}D_{LS}} \over {D_{S}}}$, and take the mean.  This mean,
$386.8\pm23.2$ Mpc, corresponds to an effective source redshift of
$1.31\pm0.04$.  Again, this is not the same as the mean source
redshift ($z=1.45$).

\begin{figure}
\centering \includegraphics[width=.70\columnwidth]{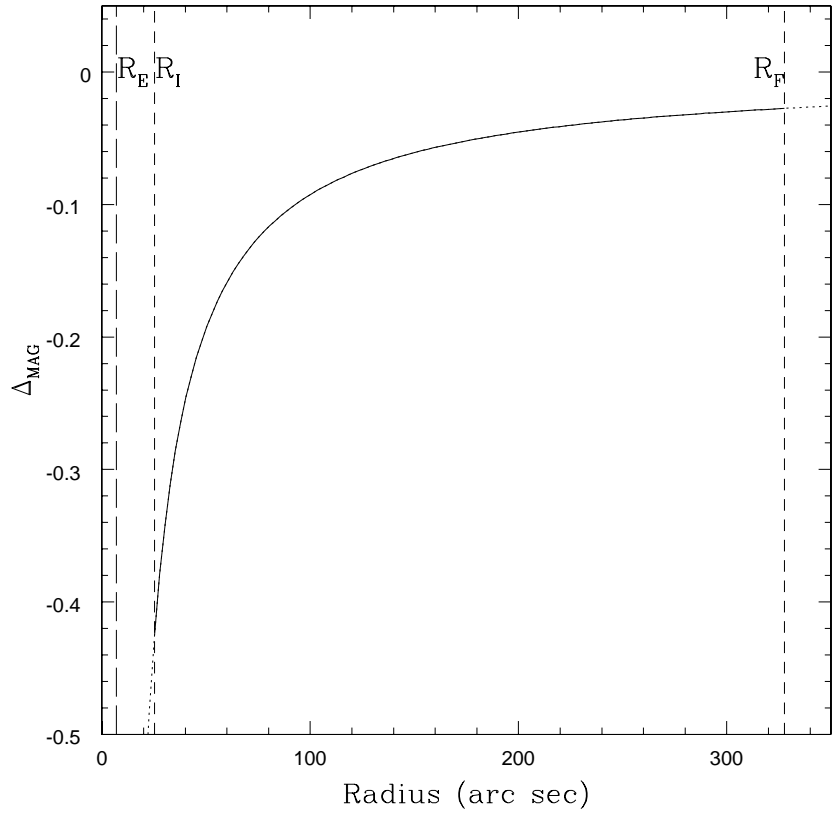}
\caption{Expected magnification of a source at $z=1.31$ by a 1004 km
s$^{-1}$ cluster at $z=0.90$, as a function of projected radius. Also
shown are the Einstein radius ($R_E$),and the inner ($R_I$) and outer
($R_F$) radii inside which the shear was computed. }
\label{fig:magnification}
\end{figure}

Finally, we check that the magnification provided by the cluster does
not significantly change the source redshift distribution near the
cluster center.  Figure~\ref{fig:magnification} shows the
magnification for a source at $z=1.31$ expected for a 1004 km s$^{-1}$
cluster at $z=0.90$, as a function of projected radius.  It is at most
0.4 mag, at the innermost radius used for tangential shear
measurement. Because of the nearly constant redshift distribution
seen in Figure~\ref{fig:magz}, this additional depth has negligible
impact on the effective source redshift.

\begin{figure}
\centering \includegraphics[width=.70\columnwidth]{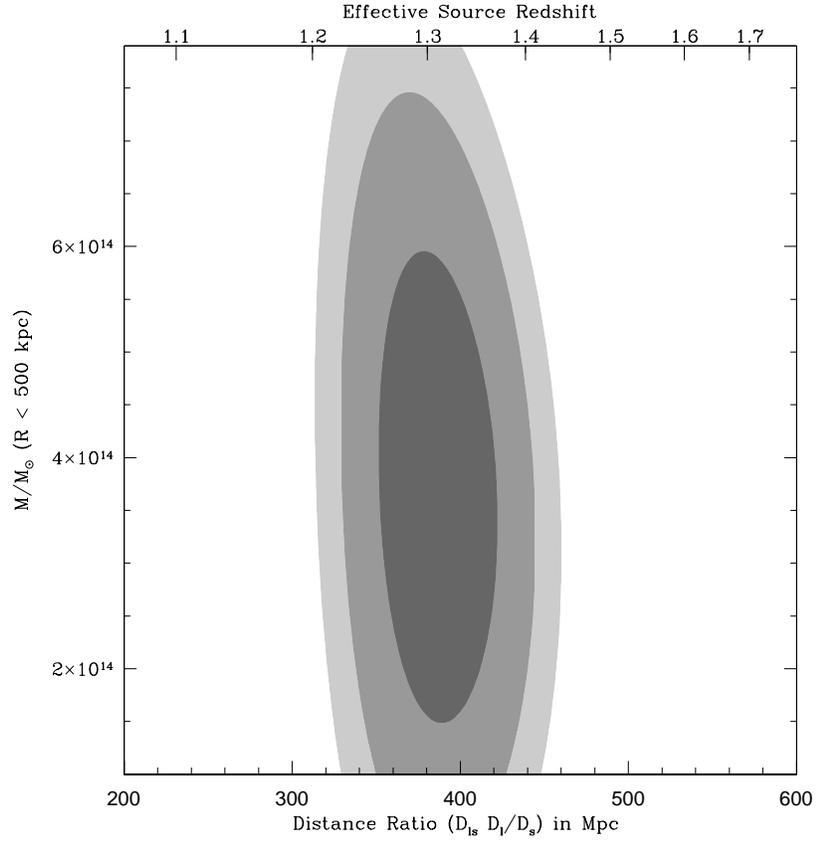}
\caption{Shear and distance ratio relative uncertainties.  The shaded
areas indicate 1, 2 and 3$\sigma$ confidence limits for the fit shown
in Figure~\ref{fig:ann}.  For reference, we also show the effective
source redshift on the top.}
\label{fig:mass}
\end{figure}

The relative uncertainties of the shear and the distance ratio
statistics are shown in Figure~\ref{fig:mass}.  The shaded areas
indicate 1, 2 and 3$\sigma$ confidence limits for the fit shown in
Figure~\ref{fig:ann}.  Combining the shear measurement with the mean
distance ratio, the weak lensing mass estimate for this cluster is
$3.69\pm1.47 \times ({R\over 500~{\rm kpc}}) 10^{14}$ $M_\odot$.  This
corresponds to an inferred velocity dispersion of $1004\pm199$ km
s$^{-1}$, which agrees with the measured velocity dispersion of
$989^{+98}_{-76}$ \citep{gal04}.  Our weak lensing measurement is also
reasonably consistent with the velocity dispersion ($\sigma_v$)
inferred from the X-ray temperature ($T_x$) of Cl 1604. Using the
best-fit to the $\sigma_v - T_x$ relation for clusters at $z < 0.5$
(Mushotzky \& Scharf 1997; Horner 2001), the measured temperature of
$2.51^{+1.05}_{-0.69}$ keV (Lubin et al.\ 2004) implies a velocity
dispersion of $575^{+110}_{-85}$ km s$^{-1}$ which is consistent
within $\sim 1.9\sigma$.  However, given the large non-statistical
scatter in the local $\sigma_v-T_x$ relation, the differences between
the velocity dispersion measurements may be well less than 2$\sigma$.

\section{Summary}

We report the weak lensing detection of Cl 1604+4304 at a redshift of
$z=0.90$. This is the highest-redshift cluster yet detected with weak
lensing.  We find a mass estimate of $3.69^{+1.47}_{-1.46} \times
({R\over 500~{\rm kpc}}) 10^{14}$ $M_\odot$, independent of the
cluster's dynamical state.  This mass estimate is in good agreement
with the spectroscopic velocity dispersion.

Unfortunately, we cannot measure the optical mass-to-light ratio (M/L)
because the small field does not provide adequate control regions.  A
wider field would serve three purposes: (1) provide control regions
for the luminosity distribution, (2) allow a direct mass
reconstruction without mass sheet degeneracy, and (3) show the larger
scale structure associated with Cl 1604+4304. The former benefit is of
great interest because we know that Cl 1604+4304 is a member of a
supercluster which contains at least four massive clusters and extends
up to $\sim 100$ Mpc \citep{gal04}. Because the dynamics of such a
large scale structure are very complex, mass measurements which do not
rely on galaxy velocities are essential. In addition to a wide field,
one must also go extremely deep in good seeing conditions to perform
an adequate weak lensing analysis of clusters at $z\sim1$.  The
combination of wide and deep, now available at a few facilities such
as the Subaru telescope in conjunction with Suprime-Cam (Miyazaki et
al.\ 2002), offers the possiblity of constraining the mass of $z\sim
1$ clusters, including Cl 1604+4304, independent of their star
formation history or dynamical state.

\acknowledgments

We would like to thank C.D. Fassnacht, R.R. Gal, J.A. Tyson and the
anonymous referee for their useful comments and contributions to this
paper. Data presented herein were obtained at the W.M. Keck
Observatory, which is operated as a scientific partnership among the
California Institute of Technology, the University of California and
the National Aeronautics and Space Administration. The Observatory was
made possible by the generous financial support of the W.M. Keck
Foundation.

\newpage


\begin{deluxetable}{ccccc}
\tablecaption{Shear as a function of radius}
\tablewidth{0pt}
\tablehead{\colhead{Radius (arc sec)} &\colhead{$\gamma_{t}$} &\colhead{$\sigma(\gamma_{t})$} &\colhead{$\gamma_{45^\circ}$} &\colhead{$\sigma(\gamma_{45^\circ})$} }
\startdata
 32   &  0.136   &   0.074  &   0.048   &   0.076   \cr
 61   &  0.076   &   0.039  &  -0.025   &   0.042   \cr
 110  &  0.009   &   0.025  &   0.037   &   0.025   \cr
 193  &  0.012   &   0.028  &   0.047   &   0.029   \cr
\enddata
\label{tab:shear}
\end{deluxetable}

\newpage


\end{document}